# Integrated ICN and CDN Slice as a Service

Ilias Benkacem[1], Miloud Bagaa[1], Tarik Taleb[1], Quang Nguyen[2], Tsuda Toshitaka[2] and Takuro Sato[2]
[1] Aalto University, Espoo, Finland
[2] Waseda University, Tokyo, Japan
Emails: [1]{firstname.lastname}@aalto.fi; [2]{firstname.lastname}@aoni.waseda.jp

*Abstract*—In this article, we leverage Network Function Virtualization (NFV) and Multi-Access Edge Computing (MEC) technologies, proposing a system which integrates ICN (Information-Centric Network) with CDN (Content Delivery Network) to provide an efficient content delivery service. The proposed system combines the dynamic CDN slicing concept with the NDN (Named Data Network) based ICN slicing concept to avoid core network congestion. A dynamic *CDN slice* is deployed to cache content at optimal locations depending on the nature of the content and the geographical distributions of potential viewers. Virtual cache servers, along with supporting virtual transcoders, are placed across a cloud belonging to multiple-administrative domains, forming a CDN slice. The *ICN slice* is, in turn, used for the regional distribution of content, leveraging the name-based access and the autonomic in-network content caching. This enables the delivery of content from nearby network nodes, avoiding the duplicate transfer of content and also ensuring shorter response times. Our experiments demonstrate that integrated ICN/CDN is better than traditional CDN in almost all aspects, including service scalability, reliability, and quality of service.

## I. INTRODUCTION

Network softwarization is gaining lots of momentum as an important concept of the fifth generation (5G) mobile systems [1]. Network slicing, leveraging Network Function Virtualization (NFV) [2], Software Defined Networking (SDN), and Multi-access Edge Computing (MEC) [3], [4] is seen as a key enabler of network softwarization. Network softwarization does not concern only the virtualization of mobile networks [5], [6], [7], but also the virtualization of service delivery platforms such as Content Delivery Networks (CDNs) [8], [9].

Information-Centric Networking (ICN), alternatively known as Content-Centric Networking (CCN), is a future Internet architecture that focuses on the content instead of the host (content publisher). ICN is based on the concept of decoupling names from locations, binding names to content, and providing a publish/subscribe model to retrieve the content by name. In-network storage is introduced in ICN to reduce the overall network traffic. Accordingly, data copies, called replicas, are dynamically created and cached within the network [10].

Content delivery, especially video content, is a major contributor to mobile traffic, occupying around 70% in mobile communications. This video traffic is expected to increase much further in case of 5G. To cope with this ever-increasing mobile video traffic, this paper proposes a new approach that aims at providing an efficient content delivery service by combining the dynamic CDN slicing concept [8], [11] with the ICN slicing concept [12], [13], [14]. In this integrated ICN/CDN slicing, ICN takes care of request routing, in-network caching and delivery of popular content [15], while CDN serves and publishes the latest content only.

First, we present an architecture which allows content providers to deploy virtual CDN and ICN slices dynamically over a federated multi-domain edge cloud. Our system exposes a northbound API over which customers can request the creation of *CDN slices* for a specific duration as a content publisher and *ICN slices* to cache the content inside the network itself and forward it back, when requested, to the interested users without reaching the CDN slice. CDN slices consist of Virtualized Network Functions (VNF) deployed over the multi-domain cloud. These VNFs include virtual transcoders, virtual streaming servers, and virtual caches, appropriately chained and configured with optimally-assigned resources to meet specific service-level performance requirements. Different methods can be used for the placement of these VNFs over the underlying cloud along with determining the resources that should be allocated for each [8], [16], [17]. ICN slices consist of ICN nodes and ICN gateways which present the intermediate component between the content/service management layer and the network layer.

The remaining of this paper is structured as follows. Section III introduces our envisioned architecture for integrating CDN slices with their respective ICN slices. The performance of the overall integration is evaluated using real testbed experiments in Section IV. Section V provides an overview of research work relevant to CDN and ICN. The paper concludes in Section VI.

## II. CDN ENHANCED WITH ICN: USE CASE

In this section, we describe the proposed framework that aims to integrate a CDN slice with ICN slice(s). The overall network system envisioned for integrating CDN and ICN slices is shown in Fig.1. In the envisioned system, a CDN slice consists of multiple virtual caches, globally deployed across multiple cloud domains and supported by multiple virtual transcoding servers to convert original video content into several standards and resolutions. A CDN slice is connected to its respective ICN slice via an ICN gateway. While a CDN slice can deliver content to viewers at the Globe scale, an ICN slice is used for the delivery of content over a specific region. The placement of CDN caches can be decided based on diverse metrics [8]. They are considered as content publishers to ICN slices. Indeed, for an efficient integration between a

CDN slice and its respective ICN slices, a gateway function is deployed in order to communicate with the CDN slice and convert the content into the ICN/NDN compatible packet format. The gateway function is provided in the ICN slice and is considered as the controller of ICN nodes of the ICN slice. From the viewpoint of traffic reduction, it is better to place the gateway functions closer to the cache servers. However, the number and the locations of caches may change according to the needs of users. Thus, there is a need to manage the number and locations of ICN gateways as per the changes happening at the CDN slice.

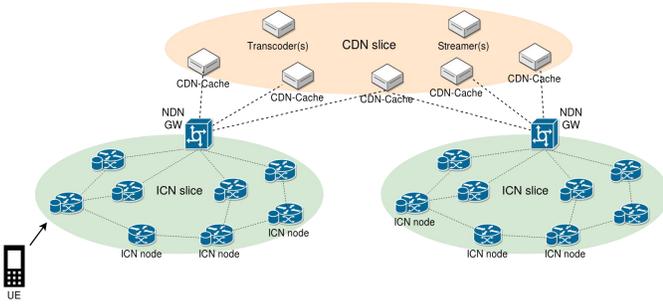

Fig. 1. Overview on the integration of a CDN slice with its respective ICN slices.

In the ICN/CDNaaS implementation, we propose Algorithm 1, whereby the subscribers (e.g., slice owners) authenticate to the multi-domain orchestrator and create CDN slices of VNF instances distributed over their envisioned service area (i.e, it can be for a specific country, continent, or at the scale of the world), then link the CDN slice to an existing ICN slice of Named Data Network (NDN) nodes through its respective NDN gateway [18], [19]. Users can upload videos to the platform, and operate the transcoding process of these videos remotely using virtual transcoders. Upon the first request for a particular content from a particular region, the requesting viewer will be served from the nearest cache in the CDN slice (e.g., `Retrieve_content_from_CDN_slice()`). Simultaneously, the content will be published into the ICN slice, serving the location of the viewer, through the NDN gateway (e.g., `Publish_Content_to_ICN()`). Subsequently, for the upcoming requests for the same content from the same region, the content will be served from the ICN network instead of being serviced from the CDN caches. Each ICN node (i.e,. virtual router + content storage) will cache any chunk of the video content traversing the network while routing back the content to the requester. Hence, the next requests will be retrieved from the ICN slice without reaching the CDN slice (e.g., `Retrieve_Content_from_ICN()`). In-network caching strategies are defined locally at the NDN nodes based on several criteria, such as expiry time and hit-ratio.

### III. ICN/CDNaaS ARCHITECTURE

The aim of the proposed architecture, depicted in Fig. 2, is to enable quick and flexible programmability to adapt to various user demands while ensuring better service delivery

**Algorithm 1** CDN enhanced with ICN.
```
.Create a CDN slice as a content
delivery service provider.
.Upload videos to CDN slice.
.Select a virtual edge transcoder.
.Create an ICN slice for request routing
and in-network caching.
.Link CDN slice(s) with the ICN slice
through a NDN-Gateway.
if First request then
  Retrieve_content_from_CDN_slice()
  Publish_Content_to_ICN()
else
  Retrieve_Content_from_ICN()
end if
```

time. It also enables ICN/CDN slicing across multiple administrative cloud domains. In the following, we describe the main components shown in Fig. 2:

*Multi-Domain Orchestrator:* is the main component of the architecture. It is responsible for managing the whole system. This component has the global view of the overall topology and represents the multi-domain orchestration plane. The CDN/ICN orchestrator allows content providers to deploy virtual CDN instances dynamically over a federated multi-domain edge cloud. Our system exposes a northbound API through which subscribers can request the creation of *CDN slices* and *ICN slices* for a specific duration over a specific area of service. A CDN slice consists of VNFs, deployed over the multi-domain cloud, including virtual transcoders, virtual streaming servers, and virtual caches. From another side, the ICN slice consists of VNFs, also deployed over the multi-domain cloud, including virtual routers and in-network caching components which forward the request hop-by-hop and routes back the data to the end-users. These VNF instances are appropriately chained together and configured with optimally-assigned resources for specific service-level performance targets. The CDN/ICN orchestrator constantly updates VNF managers regarding any action made to their respective VNFs under their directions.

*Slice-specific VNF Manager:* is considered as the brain of the target slice. The slice-specific VNF Manager communicates with the *CDN manager* and *ICN manager* reporting the virtual resources usage at the nodes level. In case of overloaded ICN or CDN slice, the respective *ICN manager* or *CDN manager* will report the needs to the VNF manager. The latter will validate the request and send it to the orchestrator to add more resources in the suitable administrative cloud domains by connecting to their respective Virtualized Infrastructure Managers (VIM). Also, the CDN/ICN managers ensure the communication between the separate groups of VNF instances (e.g., CDN caches, transcoders, streamers, and NDN nodes).

*Domain-specific VIM:* As described by the ETSI NFV-MANO standards, the VIM is responsible for carrying out re-source management and allocating virtual resources to VNFs. Virtual resources include computation resources, storage and networking resources. The NFV orchestrator communicates with VIM(s) through a Southern API to respond to the needs of slices for virtual resources.

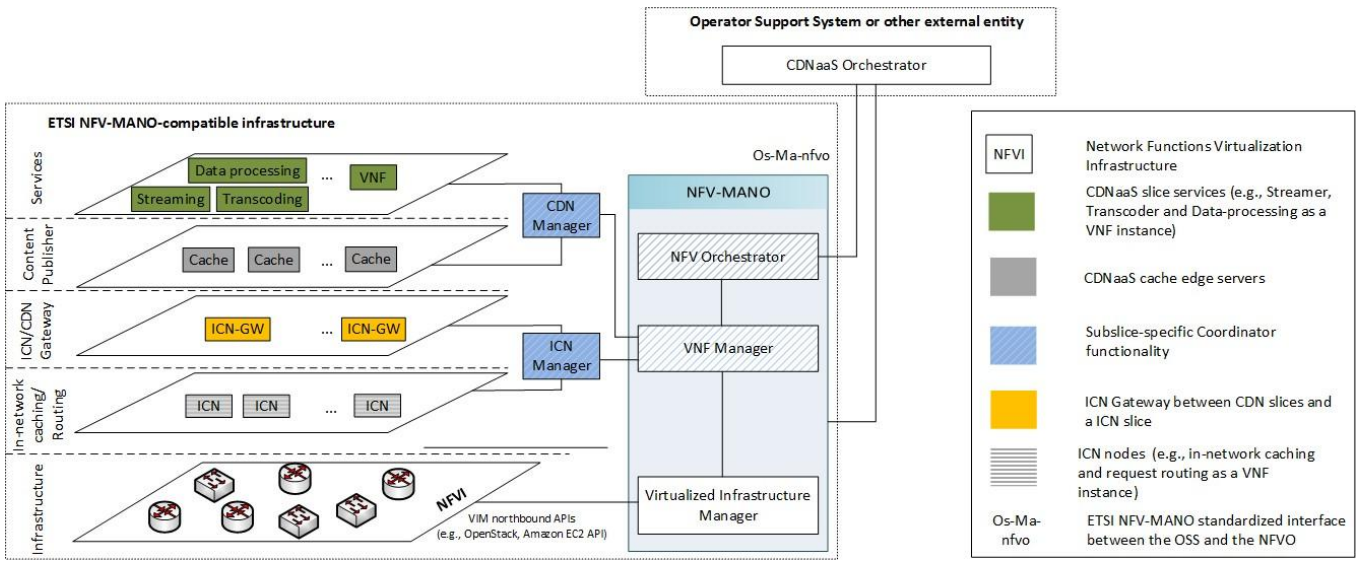

Fig. 2. Our design provides the support and the appropriate building blocks to collect and utilize different information, and it does so in a standards-based way. Our envisioned architecture complies with the ETSI NFV Management and Orchestration (NFV-MANO) framework [20].

*CDN-VNFs:* consist of three main functions: (*i*) Cache nodes that store the content published by the end-users. It is considered as the Content Publisher (CP) from the point of view of the ICN slice, (*ii*) Transcoder nodes are considered as a virtual network function that transcodes remotely videos from virtual cache servers to make the contents available in different qualities and resolutions for the users and (*iii*) Streamer nodes are considered as a virtual network function for load balancing end-user requests and streaming the selected video with an appropriate resolution.

*Dynamic NDN Gateway:* is assigned dynamically at each ICN slice based on distance, either to the content publishers or the end user distribution. The ICN node selected as gateway should have a considerable amount of resources due to the heavy work it performs while converting and publishing the content from the CDN slice to the ICN slice. The NDN gateway has the protocol translation function between ICN and IP when needed and reads out the content cached in CDN server then reformats the content and transmits it chunk-by-chunk to the next NDN node or end-user.

*NDN node (Routing/in-network caching):* is a network node with full function of ICN protocol so that the forwarding, caching functions and data exchange with content repository are handled in a manner that satisfies the network resources and the underlying network configurations and policies. The NDN node utilizes PIT (Pending Interest Table) and FIB (Forward Information Base) as the fundamental data structures for forwarding process and Content Store (CS) in order to cache the content in the network. ICN nodes are equipped with the function to aggregate requests to the same content objects to reduce network traffic and server load. NDN node components consist of three main parts: (*i*) PIT: records retrieval path and aggregates requests of the same content. It indicates the return path of the requested contents and requests aggregation function when the same content request arrives at the node; (*ii*) FIB: forwards requests based on content name: Name-based forwarding; and (*iii*) CS: acts as in-network cache storage of the ICN node to reduce the duplicated traffic of the same content objects, and also shorten the response time.

*ICN Management function:* is responsible for network management functions. It handles the key network management functions, such as configuration, network performance, QoS, congestion control, and fault tolerance management.

*Security function:* ensures an efficient secure mechanism that includes the availability, authentication and integrity functions. ICN is equipped with a mechanism to examine and confirm the authenticity of consumers, and that content object is accessible only by the authorized consumers. It controls the access and enables as well the *network security* function from malicious attacks. Each ICN node provides a mechanism to ensure that the content objects published in the network are *available* for authorized consumers and it is equipped with a tool to examine and confirm the authenticity and *integrity* of content objects.

In summary, this architecture proposed for the integration of CDN with ICN can demonstrate a solution for one of the 5G network problems on handling the increasing contents traffic. It provides an efficient content delivery service covering a wide geographical area that extends across the continents by the strength of pre-planned resource allocation [8] with various services that CDN slice will provide. It also provides less volume of the regional traffic and shorter response time of the ICN slice by the use of in-network caching capacity closer to end-users. The performance evaluation of this architecture is carried out in the following section IV.

## IV. PERFORMANCE EVALUATION

In this section, we carry out a testbed experiment to evaluate the performance of the the proposed framework that integrates CDNaaS with ICN. First, we focus on the mechanisms to publish content from CDN to ICN slice. Then, we present experimental results measuring the performance of content delivery through an integrated ICN/CDN slice.

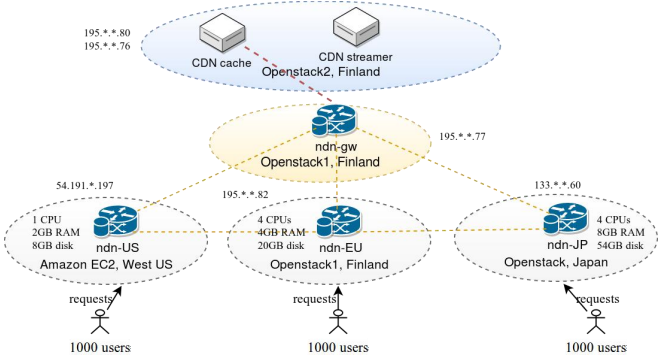

To evaluate the performance of the whole integration, we set up our testbed using four IaaS cloud platforms: Two different OpenStack clouds in Finland, $1$ OpenStack cloud in Japan and $1$ Amazon EC2 in US West (Oregon). During the experiment, we consider a CDN slice that consists of CDN cache and streamer deployed on Openstack1 in Finland. We have also used an ICN slice that includes three NDN nodes distributed over three continents and hosted on Amazon EC2 in the West US, OpenStack in Japan and our second OpenStack in Finland. Finally, an NDN/CDN gateway is deployed in Finland, as illustrated in Fig.IV. These VNF instances were created from a Ubuntu server cloud image using a customized flavor according to Table I. To introduce high load in the VM hosting the video service (i.e., busy CPU and exhausted RAM), parallel concurrent connections towards the server were launched with a total of 1000 requests being sent to the slice.

TABLE I
TESTBED SETUP

| Node | Type | Cloud | Region | Specs |
|---|---|---|---|---|
| ndn-JP | ICN | Openstack | Japan (JP) | 4CPUs, RAM 8GB, Disk 54GB |
| ndn-EU | ICN | Openstack | Finland (EU) | 4CPUs, RAM 4GB, Disk 20GB |
| ndn-US | ICN | Amazon AWS | Oregon (US) | 1CPU, RAM 2GB, Disk 8GB |
| ndn-gw | ICN Gateway | Openstack | Finland (EU) | 4CPUs, RAM 4GB, Disk 20GB |
| CDN | Cache | Openstack | Finland (EU) | 4 CPUs, RAM 4GB, Disk 120GB |

In the beginning, the CDN cache contained a set of videos, linked to a NGINX-based streaming server for content delivery. The NDN repository is empty, which means that the ICN network does not cache any content when the experiment starts.

- **Use case scenario 1:** Contents reside only in the CDN slice. When the first request arrives, it will be forwarded hop-by-hop from a router to another till it reaches the CDN cache. Then, the NDN Gateway runs the protocol translation function between ICN and IP. Also, the NDN Gateway reads out the content cached in CDN server, reformats the content and transmits it chunk-by-chunk either to the user or to the next NDN node. Fig.3.(a) shows the time to publish content from CDN to ICN by varying the video content size. We used the NDN gateway shown in Table I.

- **Use case scenario 2:** We consider a total of $3000$ requests distributed equally over the United State, Europe, and Japan. Users are requesting the same $2MB$ video content. As explained previously, only the first request will reach the CDN cache and then the ICN slice(s) will take care of the content delivery after that. The ICN slice works efficiently for content delivery thanks to the name based access and the autonomic content caching by the network nodes. The latter enables the content delivery from the nearby network node. Thus, it will reduce the duplicated content transfer on the transmission links, and also will provide shorter response times. We ran the experiments using the setup in Fig.IV taking variant measurements including mean delivery time by a node, throughput, network load at the core and edge network, and virtual resource usage.

### A. Delivery time

The aim of our ICN/CDN integration is mainly to reduce the delivery time by caching the content in virtual routers near to the users. We run the experiment where ndn-JP is considered as the NDN node that serves the $1000$ requests coming from Japan as shown in Fig. 3.(b), and ndn-EU is considered as the NDN node that serves the $1000$ requests coming from Europe, and ndn-US is considered as the NDN node that serves the $1000$ requests coming from the United States. The first observation that we can draw from the figure is that time has a positive impact on the content delivery time. This can be explained as follows. In the beginning, the contents are served from CDN, then, along with time, users retrieve the content from the nearest virtual router (NDN node) in their respective regions. Consequently, the delivery time is reduced. In a classic CDN use case, the users from EU should benefit from the CDN location and have a shorter response time. However, Fig.3.(b) shows that users from Japan had a better delivery time thanks to the ndn-JP deployed on OpenStack Waseda University in Tokyo. In summary, the experimental results demonstrate no matter the location of the original content publisher, the data packets are stored near to end users at the routers level.

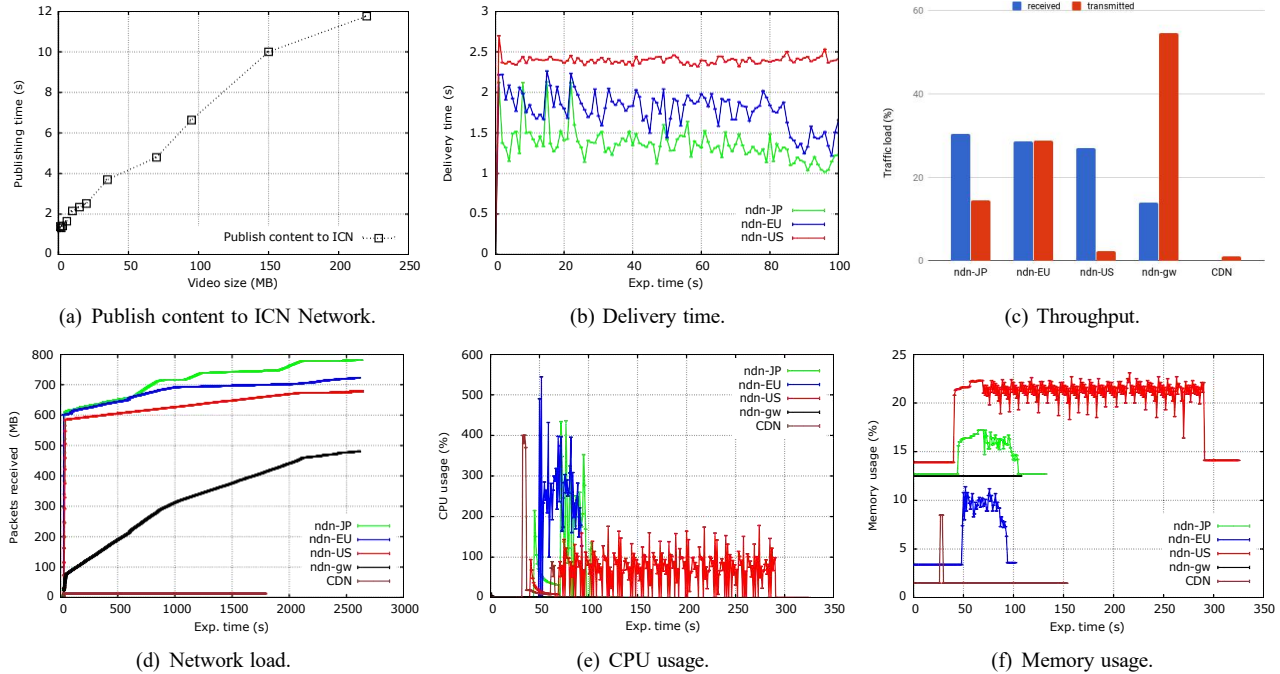

Fig. 3. Comparison of IP-based content delivery against ICN-based content delivery. We run the experiment considering 100 requests for a 2MB video content.

## B. Throughput

During the experiments, we track the input/output of all nodes using Python. Fig.3.(c) summarizes the total received and transmitted packets during the experiments. Firstly, we notice that the CDN throughput is almost nil since only the first requests reach the server. However, NDN gateway is considered as the most involved node for the transmission of packets in comparison to NDN nodes. Based on the observation that the NDN gateway is the controller node, most of the traffic pass through it. In the beginning, the content is converted from the CDN cache to the NDN gateway, then all NDN nodes retrieve the content from the gateway until the content is fully cached in the concerned NDN nodes.

The figure also indicate that the nodes (JP,EU,US) received nearly the same traffic. However, the total transmitted packets differs from a node to another which is explained by the inter-nodes communication and data exchange between NDN nodes. This demonstrate the key feature of ICN, whereby the network node with full functions of ICN protocol, including the forwarding, caching and data exchange with content repository, are handled in a manner that satisfies network resources, configurations, and policies.

## C. Network load

Fig.3.(d) supports the previous findings. At the first stage of the experiment, when an end-user requests the video content from Japan, the request will be redirected to the ndn-JP node, which is the nearest edge cloud to that user. In case that the content is not retrieved, then the node will request the next ndn-node (e.g., ndn-US node). If the content is found, then it will be routed back to the requester following the same path. In case that the requested content is not found in the NDN nodes, the request will be redirected to the NDN gateway. Moreover, the gateway node will receive the lower network traffic load since its primary function is controlling and publishing the content to the ICN network only when needed. In the reality, enhancing ICN with CDN is more efficient if we have a bigger slice of NDN nodes. Thus, either the NDN-GW and CP will be barely reached.

## D. Virtual resource usage

In our developed ICN/CDNaaS platform, we use a default in-caching strategies at the NDN node. We keep track of some measurements regarding the network, virtual resources usage (e.g., CPU and memory) to detect the involved nodes during a single request. In both Figs. 3.(e) and (f), the CDN uses the least virtual resources. The resources are used only in the beginning of the experiment. The duration of active usage of virtual resources differs from a node to another. Moreover, the figure indicates that the lengths of CPU and memory usage in ndn-US are due to its limited resources concerning the number of cores and RAM compared to other nodes.

## V. RELATED WORK

### A. Content Delivery Network

The goal of CDN is to solve a fundamental challenge for the Internet. The idea is how to distribute and retrieve content effectively [21] while reducing the service delivery time and load on end-hosts [22]. The critical component of a CDN is the method used for achieving load balancing using request routing/redirection. According to *i*) network flow at each node, *ii*) load condition, *iii*) the distance between users and source sites, and *iv*) the response time, CDN should manage to push the content closer to end-users. This strategy reduces the user access time and mitigates the network congestion. As a result, content can be cached and redistributed by utilizing CDN caching proxies [23].

*B. Information-Centric Network*

NDN [24] or Content-Centric Networking [25] is one of the most promising future Internet architecture. In NDN architecture, the contents are located thanks to their names instead of their source and destination addresses. Basically, there are two types of packets in NDN architecture, which are interest and data. Users send the interest request that contains the name of the data they want to retrieve. NDN routers forward the interest requests using their names, which is called name-based routing and create Pending Interest Table (PIT) entries to record the incoming face and outcoming face. If the Interest meets its corresponding data, the data follows the path marked by PIT entries back to the originator of the interest request. Also, NDN contains Content Store (CS) to cache data. NDN decouples requests and responses in both time and space: both data requester and producer, each does not need to know the locations of the other. Furthermore, they do not need to be online at the same time [26]. Cheng Yi et al. suggest an in-depth analysis on forwarding of NDN [27]. In the proposed solution, when the forwarding Interest packet is forwarded, NDN routers maintain the state of every pending interest. Thus, routers are aware of the network state by observing the two-way traffic. In this case, the router could score different paths at the granularity of the name prefix according to historical data. Furthermore, immediate routers can explore multiple alternative routes to prevent the network congestion, as well as the node or link failures [12], [13], [14].

*C. CDN as a Service Slicing Over a Multi-Domain Edge Cloud*

As part of our prior research work [11], [8], [28], we presented an architecture for the provision of video Content Delivery Network (CDN) [9] functionality as a service over a multi-domain cloud. We introduced the concept of the CDN slice [1], which is the CDN service instance that is created and autonomously managed on top of multiple potentially heterogeneous edge cloud infrastructures. The content provider can handle many CDN slices running in various cloud domains. A virtual CDN slice consists of four main VNFs, namely virtual transcoders [29], virtual streamers, virtual caches, and the CDN-slice-specific coordinator for the management of the slice resources and video contents across different private and public infrastructure as a service (IaaS) providers. A Multi-Domain Orchestration manages the life-cycle of VNFs and the scaling out and in of CDN slices [30]. Our design is tailored to a 5G mobile network context, building on its inherent programmability, management flexibility, and the availability of cloud resources at the mobile edge level, thus close to the end users.

VI. CONCLUSION

In this article, a framework for integrating ICN with CDN slices is proposed aiming at providing an efficient content delivery service. In the proposed framework, a CDN slice consists of virtual cache servers placed optimally over different cloud domains. An ICN slice could be associated with a CDN slice and is used for the regional distribution of the CDN content, leveraging the name-based access approach and the autonomic in-network content caching. This ICN and CND integration allows shorter service delivery times and reduces traffic load over the entire network by avoiding the duplicate transfer of content, above all over lengthy communication links. These benefits are evaluated and validated by experiments involving three continents, namely Europe, Asia, and America represented by Finland, Japan and USA, respectively.

ACKNOWLEDGEMENT

This work was partially funded by the European Union's Horizon 2020 research and innovation program under the 5G!Pagoda project with grant agreement No. 723172. The work is also partially supported by Waseda University Grant for Special Research Projects under grant number 2018S-082.